\documentclass[lnbip]{svmultln}
\usepackage{makeidx}  
\usepackage{orcidlink}
\usepackage{cite}
\begin{document}
\mainmatter              
\title{AI and Agile Software Development: 
\\From Frustration to Success 
\\XP2025 Workshop Summary}
\titlerunning{AI and Agile Software Development - Workshop summary}  
%
\author{Tomas Herda\orcidlink{0009-0005-2912-380X}\inst{1} \and Victoria Pichler\orcidlink{0009-0006-4406-127X}\inst{2} \and
Zheying Zhang\orcidlink{0000-0002-6205-4210}\inst{3} \and \newline Pekka Abrahamsson\orcidlink{0000-0002-4360-2226}\inst{3} \and Geir K. Hanssen\orcidlink{0000-0003-2718-6637}\inst{4}}
\authorrunning{Tomas Herda et al.}   
%
\tocauthor{Tomas Herda, Victoria Pichler, Zheying Zhang, Pekka Abrahamsson, Geir K. Hanssen}
\institute{AI Center of Excellence - Austrian Post, Vienna, Austria,\\
\and
Digital Logistics Platform - Group-IT - Austrian Post, Vienna, Austria
\and
Tampere University, Tampere, Finland
\and
SINTEF, 7034 Trondheim, Norway
}

\maketitle              

\begin{abstract}        
The full-day workshop on AI and Agile at XP 2025 convened a diverse group of researchers and industry practitioners to address the practical challenges and opportunities of integrating Artificial Intelligence into Agile software development. Through interactive sessions, participants identified shared frustrations related to integrating AI into Agile Software Development practices, including challenges with tooling, governance, data quality, and critical skill gaps. These challenges were systematically prioritized and analyzed to uncover root causes. The workshop culminated in the collaborative development of a research roadmap that pinpoints actionable directions for future work, including both immediate solutions and ambitious long-term goals. The key outcome is a structured agenda designed to foster joint industry-academic efforts to move from identified frustrations to successful implementation.
\keywords {Artificial Intelligence, AI and Agile, Agile Software Development, Research Roadmap, Human-AI Collaboration, Prompt Engineering, AI Governance, Software Engineering, Workshop Report}
\end{abstract}
\section{Workshop overview}
The "AI and Agile Software Development: From Frustration to Success" workshop at XP 2025 brought together a diverse group of researchers and industry practitioners. The event was designed to move beyond theoretical discussions and address the practical realities of integrating Artificial Intelligence into Agile software development. 
While numerous studies such as \cite{cinkusz2024towards, bahi2024integrating, zhang2024llm, manish2024autonomous, cabrero2024exploring} have explored the use of AI in Agile contexts, industrial experience and practitioner feedback remain underrepresented yet essential to grounding and advancing this research. To bridge this gap, the workshop deliberately invited contributions from both academia and industry. It marked the third consecutive year of a dedicated AI and Agile event at the XP Conference, building on the foundations of previous workshops at XP2023 and also AI and Agile Industry and Practice Track at XP2024\footnote{\scriptsize
XP2024\href{https://agilealliance.org/xp2024/industry-and-practice-ai-and-agile/}{AI and Agile - Industry and Practice Track}}. The significant interest in the topic was clear from the 17 total submissions received, from which two keynotes, three peer-reviewed research papers, and three industrial experience talks were accepted for presentation. The workshop successfully brought together a diverse group of 35 industrial practitioners and researchers.
\newline
\newline
The workshop had four primary goals:
\begin{itemize}
    \item To explore the intersection of AI and Agile methodologies.
    \item To share real-world experiences, including both challenges and successes.
    \item To collaboratively build future research pathways for the industry.
    \item To use AI tools for preparation, participant support, and post-event access.
\end{itemize}

The organizing committee represented a balanced academic-industrial consortium, comprising practitioners from Austrian Post and researchers from Tampere University and SINTEF. 
The workshop format guided participants from sharing individual experiences toward a collective understanding, culminating the co-creation of a research roadmap.

 A noteworthy innovation was the deliberate integration of AI-generated outputs 
 to document and extend the workshop activities. This resulted in several digital artifacts designed to capture and share the event's outcomes. An AI-generated song captured the collaborative spirit, while a custom GPT was developed to serve as a persistent, interactive knowledge base for anyone interested in the workshop's content. These resources\footnote{\scriptsize
Final Workshop Program: \href{https://conf.researchr.org/home/xp-2025/aiandagile-2025\#program}{https://conf.researchr.org/home/xp-2025/aiandagile-2025} \newline
Official Workshop Website: \href{https://gpt-lab.eu/ai-agile-workshop-xp2025/}{gpt-lab.eu} \newline
Custom Conference GPT: \href{https://chatgpt.com/g/g-68341f9d6c8c8191847064ae11c08098-aiandagilexp2025gpt}{chatgpt.com} \newline
AI-Generated Workshop Song: \href{https://suno.com/song/3123f552-9b00-486b-a552-87ad01bbbdac}{suno.com}\newline
Workshop Opening Talk: \url{https://youtu.be/xigzwCzttV4}\newline
Workshop Highlights Video: \url{https://www.youtube.com/watch?v=TmQwxuRAOIk}
}, along with video recordings and the official program and workshop website, provide a comprehensive and lasting record of the discussions.

To demonstrate the practical application of the workshop's themes, the organizers used a variety of AI tools throughout the entire workshop lifecycle.
\begin{itemize}
    \item \textbf{Creative Content \& Preparation:} The official workshop theme song was generated by \textbf{Suno.com} using custom instructions, with lyrics written by the \textbf{GPT-o4-mini} model. The presentation slides used throughout the day were created with the \textbf{Gamma.app AI Tool}, featuring images generated by the \textbf{Flux Fast 1.1} model.

    \item \textbf{Data Processing \& Analysis:} After the retrospective session, handwritten notes on flipcharts were transcribed using the \textbf{Gemma 3 27B} model. The collected frustrations were then systematically grouped and categorized with the help of the \textbf{GPT-4o} model.

    \item \textbf{Reporting \& Knowledge Sharing:} An initial draft of this workshop summary was created with assistance from the \textbf{Gemini 2.5 Pro} model. Furthermore, the \textbf{AIAndAgileXP2025GPT}, a custom knowledge base, was created using OpenAI's platform. It runs on the \textbf{GPT-4o} model and was given custom instructions to ``Act as a friendly, knowledgeable assistant...'' along with access to all accepted papers, keynotes, and this summary, creating a lasting knowledge base accessible to anyone long after the workshop has concluded.
\end{itemize}

The remainder of this summary follows the chronological structure of the workshop agenda. Section 2 describes the opening networking activities and keynote presentations that framed the day. Section 3 captures the retrospective session, where participants articulated frustrations, successes, and key lessons from AI integration in Agile contexts. Section 4 presents the review and ideation talks, highlighting current practices and future directions. Section 5 synthesizes the collaborative research roadmap developed during the final session. The summary concludes in Section 6 with a call to action and the proposal for a Living Lab format to continue advancing this emerging research-practice agenda.


\section{Setting the Stage: Networking and Keynotes}
The day began with a facilitated networking session designed to connect peers and establish a collaborative atmosphere. The session was structured in two rounds to encourage both practical and creative engagement. In the first round, "Experience Sharing," participants were asked: "What's one specific challenge or success you've seen when integrating AI into Agile workflows? How did you overcome the challenge or how did you succeed?" This question immediately grounded the workshop in practical experience. The second round, "Creative Reflection," shifted the tone by asking: "'If I Were an AI…' what would your Superpower be?" This activity fostered a more imaginative mindset and built rapport among attendees before the deeper technical sessions.

Two keynotes provided foundational perspectives that framed the day’s discussions and offered diverse viewpoints on the interplay between AI technologies and Agile practices.

\subsection*{Keynote 1: eXtreme Programming with Artificial Intelligence by Joshua Kerievsky}

Joshua Kerievsky explored the relationship between AI and eXtreme Programming (XP). He argued that to craft excellent software with AI, teams get better results by following XP’s core values and practices. The talk focused on how and when XP principles help produce successful outcomes in a human-AI collaborative model. The key takeaway was that established software engineering disciplines are not obsolete in the age of AI. Instead, they provide the necessary structure and quality control to effectively guide and validate AI-generated outputs.

\subsection*{Keynote 2: Lessons learned building an AI driven Project Management Platform by Alex Polyakov}

Alex Polyakov discussed the development of ProjectSimple.ai, an AI-powered project management platform built to tackle common challenges like misaligned goals, fragmented data, and unclear project status. He introduced a four-domain model 'Behavioral, Systematic, Analytical, Adaptive' to highlight where AI can meaningfully support Agile workflows. Rather than replicating Agile frameworks, he emphasized that tools should address real team issues and provide insights that improve decision making, clarity, and adaptability. Ultimately, it is the ability to influence the right behaviors that makes the real difference.

\section{Uncovering Collective Experience: The Retrospective Session}
The first major interactive component of the workshop was a retrospective session designed to gather empirical insights from participants’ experiences with AI integration in Agile software development. Through structured small-group discussions, attendees explored three thematic areas, and they are frustrations, successes, and lessons learned. Following the small-group discussions, a Gallery Walk facilitation technique was used. This allowed all participants to circulate and review the key findings documented at each table, ensuring a shared understanding of the collective insights before the synthesis. This process revealed a remarkable consistency in the challenges, successes, and lessons learned by practitioners and researchers alike. 

  
  

\subsection*{Shared Frustrations}
 The first round focused on challenges, asking participants \textit{"What has been your biggest frustration when trying to integrate AI into your Agile workflows?"}. The frustrations identified by the groups clustered around three key themes.

\textbf{Tooling and Model Behavior:} A primary source of frustration was the tooling itself. Participants noted there were \textit{"too many tools"} to choose from, making selection difficult. They also cited a \textit{"lack of temperature controls on LLM UI"} and the problem of being \textit{"forced to use MS Copilot."} The rapid pace of change in models was a challenge, as were \textit{"slow local language models."} A significant pain point was the unreliability of AI outputs, with participants mentioning \textit{"hallucinations,"} \textit{"making same mistakes again even after fixing before,"} and getting \textit{"non-deterministic answers."}

\textbf{Data, Privacy, and Governance:} Data was a major concern. Issues included \textit{"poor data quality,"} the difficulty of controlling privacy, and a lack of clarity around \textit{"regulation compliance."} A specific and potent frustration was the opaqueness of data usage policies, captured by the statement: \textit{"We don’t know what’s behind the checkbox"} when opting out of data collection for model training.

\textbf{Human and Process Factors:} Integrating AI into teams and workflows presented its own set of problems. Participants observed an \textit{"over-reliance on AI for junior developers"} and \textit{"poor architectural choices by AI."} A critical frustration was the feeling that the \textit{"invested time"} did not always lead to \textit{"valuable outcomes."} Teams also struggled with the \textit{"early give up"} phenomenon when initial experiments with AI tools did not succeed.

\subsection*{Celebrated Successes}
Despite the frustrations, participants shared many positive outcomes. The second round shifted to positive outcomes, with the guiding question: \textit{"Where have you seen AI add real value in Agile practices, including successful use cases, impactful tools, or innovative approaches where AI has enhanced productivity or team collaboration?"} In response to the question, participants identified the following contributions.

\textbf{Productivity and Acceleration:} AI was widely successful in accelerating tasks. Examples included \textit{"quick proof of concept (PoC) for software startup,"} \textit{"headstart for code,"} and \textit{"unit test generation."} It was also valuable for \textit{"writing user stories \& acceptance tests"} and \textit{"workshop preparation."} The sentiment of \textit{"never having to start from a blank page again"} was a powerful, recurring theme.

\textbf{Content and Code Generation:} The most cited successes involved content creation. This included \textit{"code generation \& debugging,"} \textit{"code documentation,"} \textit{"creative writing,"} and using AI as a \textit{"personal writing coach."}

\textbf{Expanding Capabilities:} AI was seen as a way to blend role boundaries and help individuals perform tasks outside their core expertise, such as \textit{"help with marketing"} or \textit{"social media marketing."} It also served as a tool for discovery, helping users \textit{"discover new features"} or act as an \textit{"AI strategy advisor."}

\subsection*{Key Lessons Learned}
The final round synthesized these experiences, prompting discussion with: \textit{"What key lessons have you learned from using AI in Agile environments? What would you repeat, avoid, or do differently next time?"} The lessons learned reflected a growing maturity in how teams approach AI.

\textbf{Human Oversight is Critical:} A dominant lesson was that human judgment remains essential. Phrases like \textit{"review generated code critically,"} \textit{"the human factor is becoming important,"} and \textit{"don’t trust AI, review output"} were common. The goal is an \textit{"embracing human-AI relationship,"} not replacement.

\textbf{Skills and Literacy are Foundational:} Participants concluded that \textit{"AI literacy is key"} and \textit{"prompting is essential."} The insight that \textit{"writing prompts is similar to writing code"} suggests a need for a more disciplined approach to interacting with AI.

\textbf{Strategic Tool Use is Necessary:} Instead of relying on a single tool, a key lesson was to \textit{"use more AI tools"} and \textit{"cross-check different models."} Participants found it helpful to \textit{"develop your own GPT"} for specific tasks and understand which tool is an \textit{"expert"} at which outputs.

\section{Sharing Current Practices: The Review and Ideation Session}
This session featured a series of short presentations from both academic and industry contributors. The presentations were divided into two parts: "Review," which focused on current applications, and "Ideation," which explored future visions. It is important to note that only the three accepted research papers will be included in the official conference proceedings.

\subsection*{Review Presentations}
The review talks highlighted practical frameworks for applying AI to specific Agile challenges.
\begin{itemize}
    \item Ayman Asad Khan's presentation, "Why Adapt RAG for Agile? Challenges, Frameworks, and the Role of Evaluator Agent" explored the challenges of using Retrieval-Augmented Generation in Agile contexts and proposed a framework that includes an "Evaluator Agent" to improve output quality.
    \item Daniel Planötscher discussed "AI-driven requirements gathering" showing how AI can assist in one of the earliest and most critical phases of the software development lifecycle.
\end{itemize}

\subsection*{Ideation Presentations}
The ideation talks looked toward the future, raising important questions and presenting new research.
\begin{itemize}
    \item Dorota Mleczko asked, "Where Are All the Agile Leaders in the AI Revolution?" pointing to a need for stronger leadership to guide AI adoption in Agile organizations.
    \item Crystal Kwok presented "AI and Teamwork in Agile Software Development: A Systematic Mapping Study" analyzing existing research on how AI impacts team dynamics.
    \item Evan Leybourn and Chris Morales spoke on "From Constraints to Capabilities: AI as a Force Multiplier" framing AI as a force multiplier that can transform team potential.
    \item Daniel Planötscher presented "NLP and GenAI in Agile Project Management: A Systematic Mapping Study" offering a comprehensive overview of research in that area.
\end{itemize}

These presentations provided concrete examples and forward-thinking concepts that fueled the subsequent roadmap discussion.

\begin{figure}
    \centering
    \includegraphics[width=0.7\linewidth]{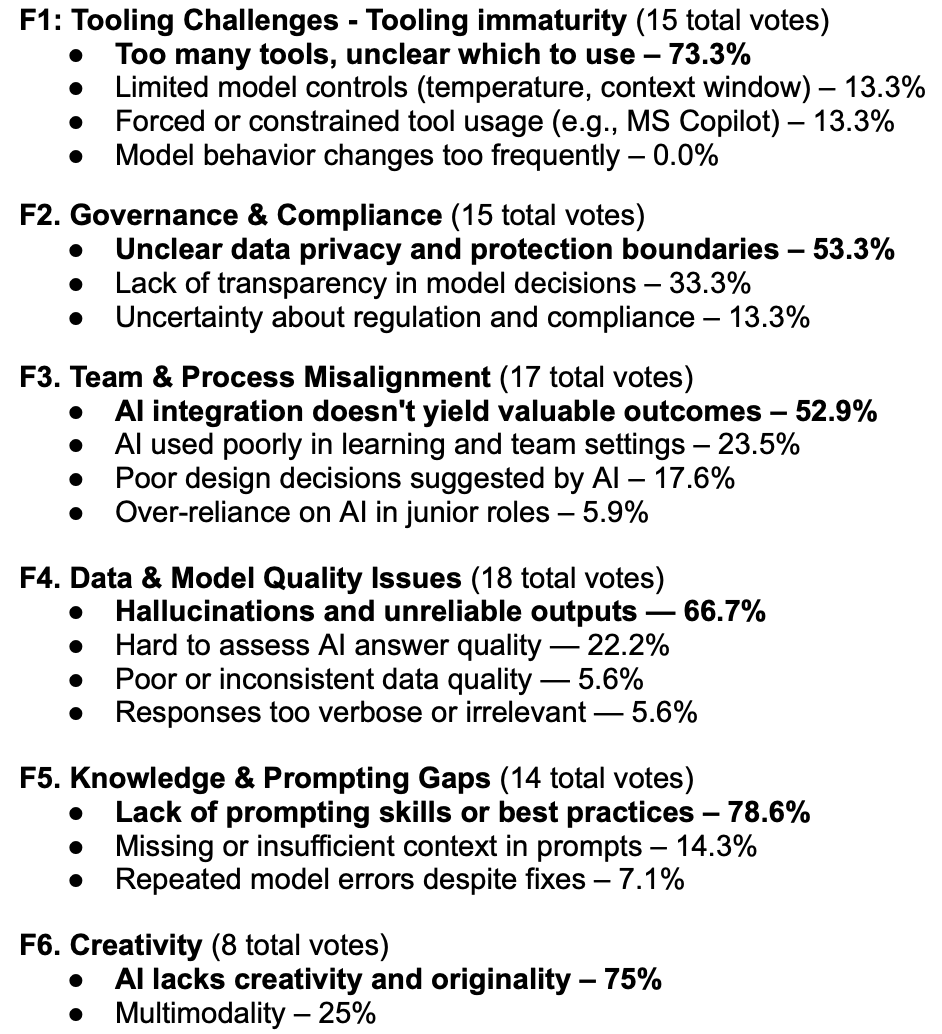}
    \caption{Voting Resutls from Participants Across Six Frustration Categories}
    \label{fig:voting results}
\end{figure}

\section{Building the Future: The Research Roadmap Session}
This final, highly interactive session was the culmination of the day's work. The objective was to synthesize the collected insights into a collaborative research roadmap.

\subsection*{Prioritizing Challenges}
The session began by presenting the six major frustration categories that were synthesized by an AI from the notes in the retrospective session. 
These categories, including tooling challenges, governance and compliance, team and process misalignment, data quality issues, knowledge gaps, and creativity limitations, provided a scaffold for focused discussion.

To prioritize areas of concern, participants voted on the most pressing sub-challenges within each category using a collaborative Padlet board \cite{Padlet_AI&Agile}. This data-driven approach ensured the subsequent deep dive focused on the issues the group collectively deemed most important. 

Figure \ref{fig:voting results} summarizes the voting results from all six frustration categories prioritized by the workshop participants. The top concerns included too many tools and lack of clarity on which to use (F1), unclear data privacy and protection boundaries (F2), AI integration that doesn't yield valuable outcomes (F3), hallucinations and unreliable outputs (F4), lack of prompting skills or best practices (F5), and lack of creativity and originality in AI outputs (F6).

\subsection*{Collaborative Deep Dive}
Participants broke into three groups, each tackling two of the prioritized frustration areas. Their task was to identify root causes, knowledge gaps, and propose both actionable short-term actions, i.e. low-hanging fruit 
and ambitious long-term 
goals i.e. moonshot ideas.

\subsubsection*{Group 1: Tooling Challenges (F1) \& Creativity Limitations (F6)}
\begin{itemize}
    \item \textbf{Root Causes:} This group identified \textit{"capitalism baked in 'new' tools"} as a driver for the overwhelming number of options. For the lack of creativity, they pointed to the inherent nature of models trained on existing data.
    \item \textbf{Knowledge \& Research Gaps:} A key gap identified was the lack of trust in AI tools and a need to better understand how to write effective prompts to elicit more creative or useful responses.
    \item \textbf{Low-Hanging Fruit:} The group proposed creating a \textit{"shared document on which tool to use for what objective"} as a practical, cross-industry resource. They also suggested focusing on the \textit{"human-AI partnership for creative use cases."}
    \item \textbf{Moonshot Idea:} Their ambitious idea was a \textit{"UI that taps into multiple tools/agents that chooses the best model"} for a given task, potentially augmented with the \textit{"thought process of people that have made successful products."}
    \item \textbf{Future Research:} Research should focus on \textit{"how to write better prompts for tools"} to improve both utility and creativity.
\end{itemize}

\subsubsection*{Group 2: Governance (F2) \& Knowledge Gaps (F5)}
\begin{itemize}
    \item \textbf{Root Causes:} Unclear privacy boundaries and a lack of prompting skills were traced back to a need for more transparency, better knowledge transfer, and more effective training.
    \item \textbf{Knowledge \& Research Gaps:} The group highlighted a need for more trust in AI, which can only be built with better data privacy, security, and specialized agents for specific tasks.
    \item \textbf{Low-Hanging Fruit:} For governance, they suggested using a \textit{"closed system, small model for testing"} to better understand AI's impact in a safe environment. To address the skills gap, they proposed a \textit{"commitment to training"} and developing a \textit{"coworker culture with AI."}
    \item \textbf{Moonshot Idea:} The governance moonshot was to \textit{"create your own LLM"} for full control. For the skills gap, they envisioned \textit{"shadow agents"} running on projects in parallel to provide comparative analysis and recommendations.
    \item \textbf{Future Research:} Research is needed on creating safe, secure environments for AI use (e.g., recording meetings) and on the effectiveness of different training methods for AI literacy.
\end{itemize}

\subsubsection*{Group 3: Process Misalignment (F3) \& Data Quality (F4)}
\begin{itemize}
    \item \textbf{Root Causes:} The group identified \textit{"missing AI literacy"} as the primary root cause for both AI integrations failing to provide value and for teams being unable to handle hallucinations. Other causes included unclear success criteria, poor data quality, and the \textit{"fear of investing in multiple LLMs."}
    \item \textbf{Knowledge \& Research Gaps:} There is a significant gap in understanding the \textit{"quantitative business benefits of AI."} Teams also lack effective strategies for validating AI output and need better \textit{"multi-LLM solutions."}
    \item \textbf{Low-Hanging Fruit:} A practical first step is for teams to clearly define success criteria and goals before implementing an AI solution.
    \item \textbf{Moonshot Idea:} The group's moonshot was to develop a system that could robustly quantify the ROI of AI integration and provide a feedback loop to fine-tune solutions.
    \item \textbf{Future Research:} Future research should focus on the \textit{"quantitative business benefits of AI,"} the \textit{"quantity and quality of AI training"} needed for teams, and the \textit{"mindsets"} that are most effective for working with AI and avoiding early abandonment of the technology.
\end{itemize}

\section{Conclusion and Call to Action}
The workshop successfully created a much-needed space for open dialogue, bridging the gap between academic theory and industry practice. The most significant outcome was the clear alignment on the core frustrations and knowledge gaps shared by both communities. The challenges of tooling, governance, data quality, and skills are not unique to one domain; they are universal hurdles on the path to integrating AI effectively into Agile workflows. This resonance with recent studies \cite{amershi2019guidelines, ji2023survey, kulkarni2017integration, liu2023pre, shneiderman2020human} indicates the widespread nature of these challenges. The day's journey, from sharing individual frustrations to collaboratively building a research roadmap, demonstrated a powerful model for progress. 

However, a roadmap is only a guide. To ensure the momentum from this workshop translates into tangible progress, we must move from discussion to action. The workshop identified that a primary reason "AI integration doesn't yield valuable outcomes" is a fundamental gap between knowing about AI and knowing how to use it effectively in a given context. Simply encouraging experimentation is not enough.

Therefore, we issue a specific call to action: \textbf{to launch the "AI and Agile Living Lab" as an interactive session at the future conference}.

This will not be a traditional track of presentations. Instead, it will be a structured, hands-on problem-solving environment designed to kickstart a continuous \textbf{``Learning Loop''} for participants and their organizations. Drawing inspiration from proven methods for accelerating AI adoption, the \textbf{Living Lab} will operate on a simple but powerful principle: \textbf{people learn best by applying AI to their own real work.}

The proposed structure for the \textbf{AI and Agile Living Lab} is as follows:

\begin{enumerate}
    \item \textbf{Bring Your Own Problem (BYOP):} Participants will come to the session with a genuine, unresolved challenge from their work at the intersection of AI and Agile. This could be a practitioner's struggle with generating test cases or a researcher's difficulty in analyzing qualitative data.

    \item \textbf{Collaborative Work in Mixed Pairs:} Participants will be paired up, intentionally mixing industry practitioners with academic researchers. This structure fosters direct collaboration and knowledge exchange.

    \item \textbf{Structured, Multi-Model Experimentation:} The lab will guide pairs to use several different AI models (e.g., GPT, Claude, Gemini) on their problem, comparing outputs and learning the unique strengths of each. This directly addresses the \textbf{``too many tools, unclear which to use''} frustration.

    \item \textbf{Guided Prompting Techniques:} Participants will be introduced to and use structured prompting methods, moving beyond simple requests to craft effective, context-rich instructions for the AI. This directly tackles the top-voted frustration: \textbf{``lack of prompting skills or best practices.''}

    \item \textbf{Focus on Tangible Outcomes and Reflection:} The goal is not just to try, but to achieve. By the end of the session, each pair will aim to have a tangible output: a working code snippet, a set of refined user stories, a draft experimental design, or a new, documented workflow. The session will conclude with pairs sharing what they built, what they learned, and the \textbf{``moment of insight''} that made AI's value concrete for them.
\end{enumerate}

By creating this \textbf{Living Lab}, we move beyond discussing the research roadmap and begin actively executing it. This format provides a direct, evidence-based response to the core challenges identified in our workshop. It builds the critical AI literacy our community needs, demonstrates tangible ROI on a personal scale, and forges the authentic practitioner-researcher collaborations required to solve our field’s most pressing problems. We invite the entire community to join us in 2026, not just to talk about the future, but to \textbf{build it together, one solved problem at a time.}

%


\bibliographystyle{splncs}
\bibliography{references}

\end{document}